\documentclass[preprint]{elsarticle}

\usepackage{lineno,hyperref}
\usepackage{amsfonts,graphicx,subfigure,bm}
\usepackage{amsmath,amssymb}
\modulolinenumbers[5]

\journal{Journal of \LaTeX\ Templates}

\begin{document}

\begin{frontmatter}

\title{Joint Frequency Estimation with Two Sub-Nyquist Sampling Sequences\tnoteref{mytitlenodte}}
\tnotetext[mytitlenote]{This work is}

\author{Shan Huang\fnref{}}
\author{Hong Sun\fnref{}}
\author{Lei Yu\fnref{}}
\author{Haijian Zhang\fnref{}}
\address{Signal Processing Laboratory, School of Electronic Information, Wuhan University, Wuhan 430072, China}

%


\begin{abstract}
In many applications of frequency estimation, the frequencies of the signals are so high that the data sampled at Nyquist rate are hard to acquire due to hardware limitation. In this paper, we propose a novel method based on subspace techniques to estimate the frequencies by using two sub-Nyquist sample sequences, provided that the two under-sampled ratios are relatively prime integers. We analyze the impact of under-sampling and expand the estimated frequencies which suffer from aliasing. Through jointing the results estimated from these two sequences, the frequencies approximate to the frequency components really contained in the signals are screened. The method requires a small quantity of hardware and calculation. Numerical results show that this method is valid and accurate at quite low sampling rates.
\end{abstract}


\end{frontmatter}


\section{Introduction}
Frequency estimation has wide applications in communications, audio, medical instrumentation and  electric systems \cite{zapateiro2014secure}\cite{liao2014analytical}\cite{arablouei2015adaptive}. Frequency estimation methods cover classical modified DFT \cite{candan2015fine}, subspace techniques such as MUSIC \cite{schmidt1986multiple} and ESPRIT \cite{roy1989esprit} and other advanced spectral estimation approaches \cite{stoica1989maximum}. In general, the sampling rate of the signal is required to be higher than twice the highest frequency (i.e. Nyquist rate). However, it is challenging to build sampling hardware when signal bandwidth is large. When the signal is sampled at sub-Nyquist rate, it leads to aliasing and attendant problem of frequency ambiguity.

A number of methods have been proposed to estimate frequency with sub-Nyquist sampling. To avoid the frequency ambiguity, Zoltowski proposed a time delay method which requires the time delay difference of the two sampling channels less than or equal to the Nyquist sampling interval \cite{zoltowski1994real}. By introducing properly chosen delay lines, and by using sparse linear prediction, the method in \cite{tufts1995digital} provided unambiguous frequency estimates using low A/D conversion rates. The authors of \cite{li2009robust} made use of Chinese Remainder Theorem (CRT) to overcome the ambiguity problem, multiple signal frequencies often need the parameter pairing. Bourdoux used the non-uniform sampling to estimate the frequency \cite{bourdoux2011sparse}. Some scholars used multi-coset sub-Nyquist sampling with different sampling rates to obtain unique signal reconstruction \cite{venkataramani2000perfect}\cite{sun2012wideband}. Based on emerging compressed sensing theory, sub-Nyquist wideband sensing algorithms and corresponding hardware were designed to estimate the power spectrum of a wideband signal \cite{tropp2010beyond}\cite{mishali2010theory}\cite{tian2012cyclic}. However, these methods usually require much hardware and complicated calculations, which makes the practicability discounted.

In this paper, we propose a new method to estimate the frequency components from under-sampled measurements. The method uses two sets of sub-Nyquist samples and requires less hardware than former methods. When the sampling rates satisfy certain conditions, the frequencies can be estimated accurately. The paper is organized as follows: Section II describes the formulation of the problem and reviews ESPRIT. Section III gives our method and analysis. Simulation results are shown in Section IV. The last section draws conclusions.
\section{Preliminaries }
\subsection{Problem Formulation}
Consider a signal containing $K$ frequency components with unknown constant amplitudes and phases, and additive noise that is assumed to be a zero-mean stationary complex white Gaussian random process. The samples of the signal at sampling rate $f_S=1/\Delta t$ can be written as
\begin{equation}\label{eq1}
 x(n) =\sum\limits_{k = 1}^K {{ s_k}{e^{j(2\pi {f_k}n\Delta t)}} + e(n)},n=1,2,\cdots,
\end{equation}
where $f_k$ is the $k$-th frequency, $s_k$ is corresponding complex amplitude, and $ e(n)$ is additive Gaussian noise. Assume that the upper limit of the frequencies is known, but we only have low-rate analog-to-digital converters whose sampling rates are far lower than Nyquist rate. Many articles use multi-channel measurement systems to estimate the frequencies. We shall demonstrate that double-channel sub-Nyquist sampling with specific rates is enough in general.
\subsection{ESPRIT}
The ESPRIT utilizes the subspace approach and estimates signal parameters via rotational invariance techniques \cite{roy1989esprit}. Define $y(n)= x(n+1)$ and choose an $m>K$. Two $m\times1$ vectors $\bm x(n)$ and $\bm y(n)$ are defined as
\begin{equation}\label{eq2}
\begin{split}
\bm x(n)&=[x(n),\cdots,x(n+m-1)]^{T},\\
\bm y(n)&=[x(n+1),\cdots,x(n+m)]^{T}.
\end{split}
\end{equation}
By using (\ref{eq1}), the time series $\bm x(n)$ and $\bm y(n)$ can be written in matrix notation, namely
\begin{equation}\label{eq3}
\begin{split}
\bm x(n) &= \bm A\bm s + \bm e(n),\\
\bm y(n) &= \bm A\bm\Phi\bm s +\bm e(n + 1),
\end{split}
\end{equation}
where $\bm s = {\left[ {{s_1}{e^{j2\pi {f_1}n\Delta t}}, \cdots ,{s_K}{e^{j2\pi {f_K}n\Delta t}}} \right]^T}$ is a $K\times1$ vector of complex amplitudes, $\bm e(n)=[e(n),\cdots,e(n+m-1)]^{T}$, and $\bm{\Phi}$ is a diagonal $K\times K$ matrix containing the relative phase between adjacent time samples for each of the $K$ components
\begin{equation}\label{eq4}
  {\bf{\Phi }} = diag\left[ {{e^{j2\pi {f_1}\Delta t}}, \cdots ,{e^{j2\pi {f_K}\Delta t}}} \right].
\end{equation}
$\mathbf{\Phi}$ is a diagonal matrix relating the temporally displaced vectors $\bm x$ and $\bm y$ and is therefore referred to as a rotation operator. $\bm A$ is the $m\times K$ Vandermonde matrix, namely
\begin{equation}\label{eq5}
{\bm{A}} = \left[ {\begin{array}{*{20}{c}}
1& \cdots &1\\
{{e^{j2\pi {f_1}\Delta t}}}& \cdots &{{e^{j2\pi {f_K}\Delta t}}}\\
 \vdots & \ddots & \vdots \\
{{e^{j2\pi {f_1}\left( {m - 1} \right)\Delta t}}}& \cdots &{{e^{j2\pi {f_K}\left( {m - 1} \right)\Delta t}}}
\end{array}} \right]
\end{equation}
The auto-covariance matrix of $\bm x$ is given by
\begin{equation}\label{eq6}
  {{\bm{R}}_{xx}} = E\left( {\bm x(n)\bm x{{(n)}^H}} \right) = \bm A\bm S{\bm A^H} + {\sigma ^2}\bm I,
\end{equation}
where $\sigma^{2}$ is the variance of $e(n)$, $\bm I$ is an identity matrix and $\bm S$ is a nonsingular matrix. $(\ast)^{H}$ denotes the Hermite transpose operation. The cross-covariance matrix of the vectors $\bm x$ and $\bm y$ is given by
\begin{equation}\label{eq8}
  {{\bm{R}}_{xy}} = E\left( {\bm x(n)\bm y{{(n)}^H}} \right) = \bm A \bm S{\bm \Phi ^H}{\bm A^H} + {\sigma ^2}\bm Z,
\end{equation}
where $\bm Z$ is a $K\times K$ matrix with ones on the first sub-diagonal and zeros elsewhere, i.e.
\begin{equation}\label{eq9}
  \bm Z = \left[ {\begin{array}{*{20}{c}}
0&0& \cdots &0\\
1&0& \cdots &0\\
 \vdots & \ddots & \ddots & \vdots \\
0&0&1&0
\end{array}} \right].
\end{equation}
$\bm R_{xx}$ has $(m - K)$ minimum eigenvalues all equal to $\sigma^{2}$. Define
\begin{equation}\label{eq10}
\begin{split}
{\bm C_{xx}} &= {\bm R_{xx}} - {\sigma ^2}\bm I = \bm {AS }{\bm A^H},\\
{\bm C_{xy}} &= {\bm R_{xy}} - {\sigma ^2}\bm Z = \bm {AS }{\bm \Phi ^H}{\bm A^H},
\end{split}
\end{equation}
and consider the matrix pencil given by
\begin{equation}\label{eq11}
  {\bm C_{xx}} - \gamma {\bm C_{xy}} = \bm {AS}\left( {\bm I - \gamma {\bm \Phi ^H}} \right){\bm A^H}.
\end{equation}

By inspection, the column space of $\bm{AS}\bm A^{H}$ and $\bm{AS}\bm \Phi^{H}\bm A^{H}$ is identical. If $\gamma=e^{j2\pi f_{k}\Delta t}$, the $k$-th row of $(\bm I - \gamma\bm\Phi^{H})$ will be zero and the pencil $(\bm C_{xx} - \gamma \bm C_{xy})$ will decrease in rank. Therefore, $e^{j2\pi f_{k}\Delta t}(k=1,\cdots,K)$ are exactly the generalized eigenvalues of the matrix pair $\{\bm C_{xx},\bm C_{xy}\}$. Once the generalized eigenvalues $\gamma$ are calculated, the frequencies of the signal can be estimated. However, when the sampling rate is lower than Nyquist rate, only a series of probable frequencies are obtained via ESPRIT. In next section, we analyze the problem and determine the real frequencies by sampling at two specific sub-Nyquist rates.
\section{Proposed Method}
\subsection{The Impact of Under-sampling}
Under-sampling usually leads to aliasing of frequency spectrum. Suppose the highest frequency component in the signal is lower than $f_{H}$, we sample at the rate $f_{S}=f_{H}/p~(p>1)$. So the samples can be written as
\begin{equation}\label{eq12}
x(n) = \sum\limits_{k = 1}^K {{s_k}{e^{j2\pi {f_k} \cdot np/{f_H}}} + e(n)} ,n = 1,2, \cdots.
\end{equation}
In the case of under-sampling, if we deal with the samples as the same with that in normal sampling, the normalized frequency $p\cdot f_k/f_H$ may be greater than 1 for some $k$. Due to the periodicity of trigonometric functions, one $e^{j2\pi {f_k} \cdot p/{f_H}}$ will correspond to multiple $f_k$.  In the following we shall expand the frequencies within $f_H$.
\subsection{The Expansion In Frequency Domain}
As a super-resolution subspace technique, ESPRIT has favourable robustness to noise. So we choose ESPRIT as the first step of our method. Then the generalized eigenvalues $\gamma$ in (\ref{eq11}) turn into $\gamma  = {e^{j2\pi {f_k} \cdot p/{f_H}}}$. Because $p \cdot {f_k}/{f_H} > 1$ for some $k$, a series of estimated frequencies $\tilde{f}_{k}$ all satisfy ${e^{j2\pi {\tilde{f}_k} \cdot p/{f_H}}}=\gamma $. We restrict the under-sampled ratio $p$ to be a positive integer. Denoting the principal argument of $\gamma$ as $Arg(\gamma)$, we obtain the estimation from sub-Nyquist samples
\begin{equation}\label{eq12a}
  {\hat{f}_k}={Arg(\gamma)\cdot f_{H}}/{(2\pi p)}, k=1,2,\cdots,K.
\end{equation}
The $\hat{f}_k$ are only a part of possible frequencies, which occupy the bottom in frequency domain. All the possible frequencies (we call eligible frequencies) can be expanded as
\begin{equation}\label{eq13}
  {\tilde f_k} = {{{\hat f}_k} + l_{k} \cdot {f_H}}/p,~l_{k} = 0,1, \cdots ,p - 1.
\end{equation}

The eligible frequencies include the frequencies approximate to the frequency components really contained in the signals (we call real frequencies) and the false frequencies caused by under-sampling. It is worth mentioning that different real frequencies may correspond to the same ${\hat{f}_k}$. This condition has no crucial influence on the following algorithms because no real frequencies will be left out from the eligible frequencies. Since this condition is not common, we assume that different real frequencies correspond to different ${\hat{f}_k}$ for the concision of discussions.
\subsection{The Screening of The Eligible Frequencies}
In order to determine the true value of $l_{k}$ for each $k$, we sample the other time series at another different rate $f'_S=f_H/q$, where $q$ is another positive integer. How to screen the real frequencies from the eligible frequencies? An intuitional method is using ESPRIT for the other sample sequence again and then matching the two series of eligible frequencies. When the number of the eligible frequencies is not large, we can match the two sets of eligible frequencies by trial and error. That is to say, we may solve proper $l_{k}$ and $l'_{k}$ from a dualistic linear indeterminate equation
\begin{equation}\label{eq14}
   {{\hat f}_k} + {l_k} \cdot {f_H} /p = {{\hat f}_k}'  + {{l'}_k} \cdot {f_H} /q,
\end{equation}
i.e.,
\begin{equation}\label{eq15}
  q{l_k} - p{l'_k} = pq\left( {{{\hat f}_k}' - {{\hat f}_k}} \right)/{f_H},
\end{equation}
where $l_{k}=0,1,\cdots,p-1$ and ${l'_k}=0,1,\cdots,q-1$, $(\ast)'$ denotes the parameter of the second sampling. According to B\'{e}zout's identity, when $p$ is coprime to $q$ (i.e. $p\perp q$), (\ref{eq15}) has integer solutions as long as the right hand side of (\ref{eq15}) is an integer. Moreover, since $l_k\in \{0,1,\cdots,p-1\}$ and ${l'_k}\in \{0,1,\cdots,q-1\}$, the equation (\ref{eq15}) just has an unique satisfactory solution. Denoting the true values of $l_{k},{l'_k}$ by $\bar{l}_{k},\bar l'_k$, according to (\ref{eq13}) we have
\begin{equation}\label{eq15c}
\begin{split}
  pq\left( {{{\hat f'}_m} - {{\hat f}_n}} \right)/{f_H} &= pq\left( {{f_m} - {f_n} + {{\bar l}_n} \cdot {f_H}/p - {{\bar l'}_m} \cdot {f_H}/q} \right)/{f_H} \\
    &=pq\left( {{f_m} - {f_n}} \right)/{f_H} + q{{\bar l}_n} - p{{\bar l'}_m}.
\end{split}
\end{equation}
To make the matching correct, (\ref{eq15}) must be an integer when and only when $m=n$. When $m\neq n$ and $\Delta f\triangleq f_m-f_n$ is the integer multiples of $f_H/\left(pq\right)$, the matching of the frequencies may be suffer from mistakes. Fortunately, in the simulations we find that the probability of this situation is quite low. The condition that under-sampled ratios $p$ and $q$ are coprime offers good properties for screening, as we will see later. We can match the eligible frequencies by graphic method. The eligible frequency points ${\tilde f_k}$ and $\tilde{f'_{k}}$ are plotted together and then the coincident points are found as the real frequencies. When there are too many eligible frequencies, the direct matching of the eligible frequencies is impractical. In the following we propose a more practical algorithm to screen the real frequencies.

We use a MUSIC-like algorithm to determine the real frequencies instead of the matching process. The second sample sequence can be written as
\begin{equation}\label{eq15a}
  x'(n)= \sum\limits_{k = 1}^K {{s_k}{e^{j2\pi qn \left({{\hat f}_k}/{f_H} + {{\bar l}_k} /p\right)}}}+ e'(n).
\end{equation}
Let $g_k={\hat f}_k/f_H$, the time series of second sampling is expressed as
\begin{equation}\label{eq15b}
  \bm x'(n) = \left[\bm a(g_{1},{\bar l_1}),\bm a(g_{2},{\bar l_2}), \cdots ,\bm a(g_{K},{\bar l_K})\right]\bm s +\bm e'(n),
\end{equation}
where
\begin{equation}\label{eq16}
 \bm a(g_k,{\bar l_k}) = \left[ e^{j2\pi q( g_k +{\bar l}_k/p )}, \cdots ,e^{j2\pi qm'( g_k +{\bar l}_k/p )} \right]^T,
\end{equation}
and $m'$ is the number of sequential samples. The subspace spanned by $\bm a(g_{k},{\bar l_k})$ can be expanded. That is to say, for each $k$, $\bm a(g_{k},{\bar l_k})$ are augmented as
\begin{equation}\label{eq17}
\bm a(g_k,\bar l_k) = {\left[ {{\bm a_0}({{\hat f}_k}),{\bm a_1}({{\hat f}_k}), \cdots ,{\bm a_{p - 1}}({{\hat f}_k})} \right]},
\end{equation}
where
\begin{equation}\label{eq18}
  \bm {a}_r(g_k) = {\left[ {e^{j2\pi q( g_k + r/p)}}, \cdots , {e^{j2\pi qm'( g_k + r/p)}}\right]^T}.
\end{equation}
So (\ref{eq15b}) changes into
\begin{equation}\label{eq19}
  \bm x'(n) = \bm A'\bm s' + \bm e'(n),
\end{equation}
where
\begin{equation}\label{eq20}
  \bm A' = \left[ {{\bm a_0}({g_1}), \cdots ,{\bm a_{p - 1}}({g_1}), \cdots ,{\bm a_0}({g_K}), \cdots ,{\bm a_{p - 1}}({g_K})} \right],
\end{equation}
and
\begin{equation}\label{eq20a}
  \bm s' = \left[ {s_1}{e^{j2\pi {f_1}n\Delta t}}, \cdots ,{s_1}{e^{j2\pi {f_1}n\Delta t}}, \cdots ,{s_K}{e^{j2\pi {f_K}n\Delta t}}, \cdots ,{s_K}{e^{j2\pi {f_K}n\Delta t}}\right] ^T.
\end{equation}

The following procedure is similar to MUSIC. Taking the eigen-decomposition of the auto-covariance matrix of $\bm x'$ renders
\begin{equation}\label{eq21}
  {\bm R_{{\bm{x'x'}}}} = \left[ {{\bm U_s},{\bm U_e}} \right]\left[ {\begin{array}{*{20}{c}}
\bm \Lambda &0\\
0&{{{\sigma '}^2}{\bm I}}
\end{array}} \right]\left[ {\begin{array}{*{20}{c}}
{\bm U_s^H}\\
{\bm U_e^H}
\end{array}} \right],
\end{equation}
where the column vectors of $\bm U_{s}$ and $\bm U_e$ are, respectively, the eigenvectors that span the signal subspace and the noise subspace of $\bm R_{\bm{x'x'}}$ with the associated eigenvalues on the diagonals of $\bm \Lambda$ and ${\sigma '}^2{\bm I}$. So the pseudo-spectrum with respect to $\tilde{f}_{k}= g_k\cdot f_H+l_k \cdot f_H/p$ can be given by
\begin{equation}\label{eq22}
{P_{MU}}({\tilde f_k}) = \frac{1}{{\bm a_r^H({g_k}){\bm U_e}\bm U_e^H{\bm a_r}({g_k})}},k=1,2,\cdots,K,r=0,1,\cdots,p-1.
\end{equation}
The maximum $K$ frequencies corresponding to the highest $K$ wave peaks or the frequencies that have the power larger than a threshold are the final estimated frequencies $\bar{f}_k$. This algorithm has less amount of computations than conventional MUSIC because the number of eligible frequencies is generally less than that of uniform grid points. In (\ref{eq20}), if $m'$ is an integral multiple of $p$, the mode vectors ${\bm a_r}({g_k})$ corresponding to the same ${\hat f}_k$ are mutually orthogonal, which makes it easy to distinguish the real frequencies. So the condition that $p$ and $q$ are required to be relatively prime provides high discriminability for screening.
\subsection{The Procedure of The Method}
\begin{figure}[!htbp]
\centering
\includegraphics[scale=0.9]{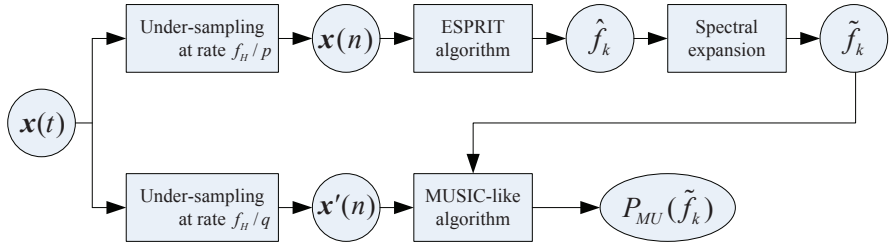}
\caption{ The procedure diagram of the method. } \label{f1}
\end{figure}

We summarize the procedure of proposed method as shown in Fig.~\ref{f1}. Firstly one sample sequence is used to obtain the eligible frequencies via ESPRIT. Then the other sample sequence is used to screen the eligible frequencies via the MUSIC-like algorithm. The maximum $K$ wave peaks indicate the real frequencies, or a proper threshold is set to filter the false frequencies.
\section{Simulation Results }
In this section, we shall create satisfactory signals and estimate the frequency components by the proposed method. The signals are composed of $K$ complex-valued sinusoids. The frequencies are assumed to be lower than $100Hz$ and the minimum interval is larger than 0.1Hz. The phase angles corresponding to the frequency components are set to be randomly distributed in $[0,2\pi)$. In the simulations, the two sub-Nyquist sequences are sampled at the rate $100/5Hz$ and $100/7Hz$ (i.e. $p=5,q=7$).

\begin{figure}[!htbp]
\centering
\subfigure[]{
\includegraphics[scale=0.6]{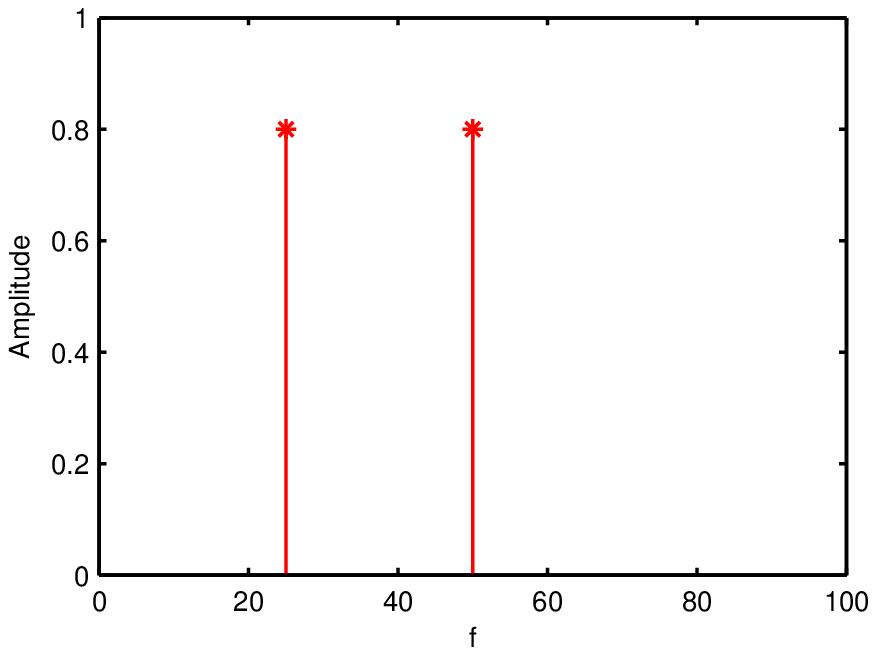}}
\subfigure[]{
\includegraphics[scale=0.6]{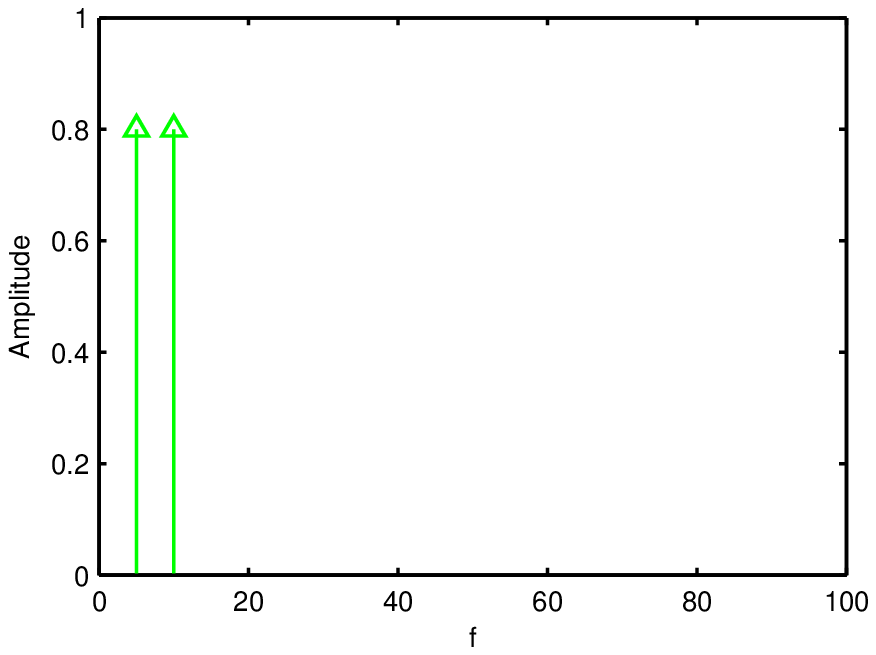}}\\
\subfigure[]{
\includegraphics[scale=0.6]{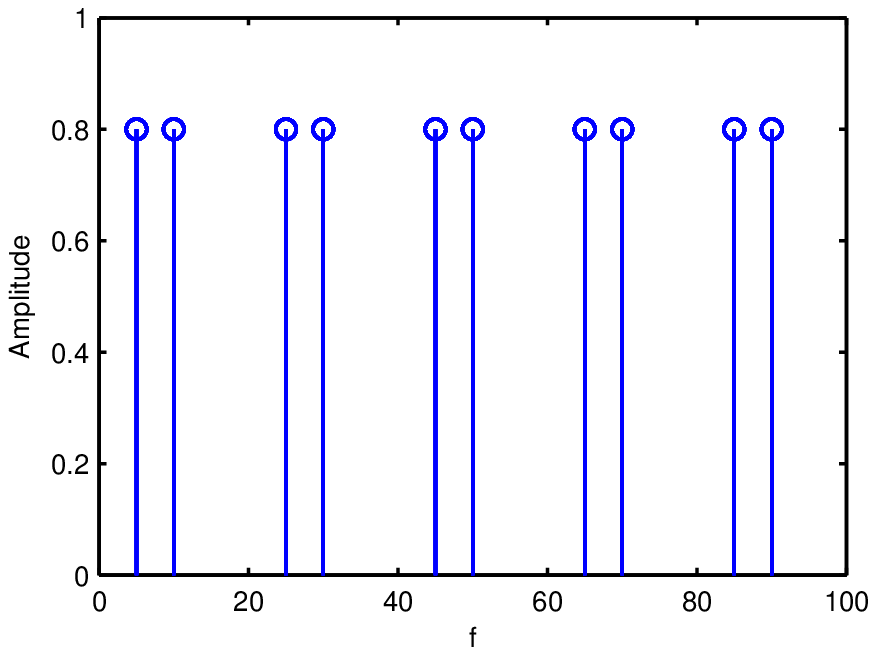}}
\subfigure[]{
\includegraphics[scale=0.6]{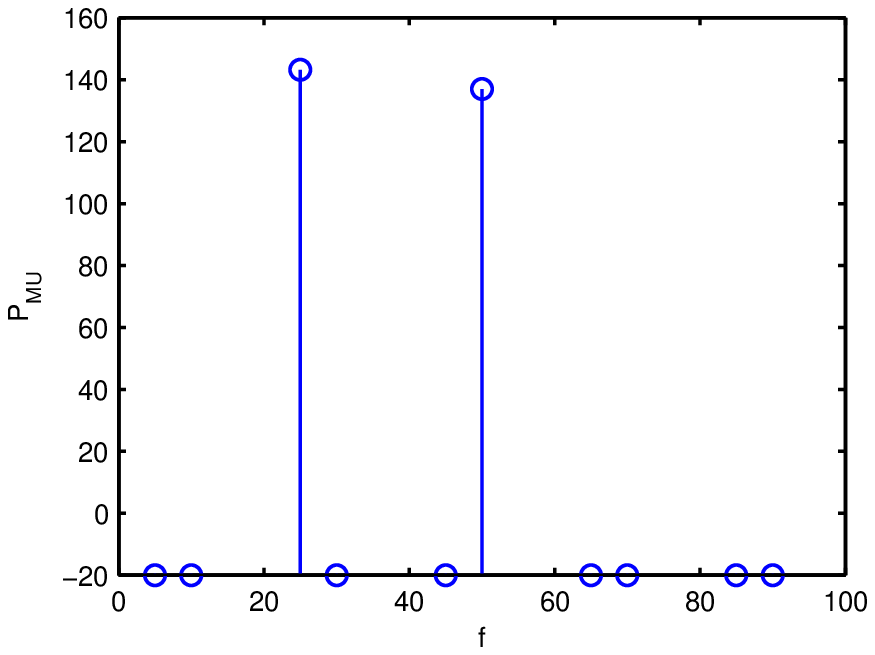}}
\caption{The comparison of the frequencies. (a) $f_k$; (b) $\hat{f}_k$; (c) $\tilde{f}_k$; (d) $\bar{f}_k$. } \label{f2}
\end{figure}

In the first simulation, we demonstrate the process of our method in noiseless condition. The number of frequency components is set to be $K=2$ and the amplitudes are 0.8. As shown in Fig.~\ref{f2}, we plot the true frequencies $f_k$, the estimated frequencies $\hat{f}_k$ via ESPRIT, the eligible frequencies $\tilde{f}_k$ and the pseudo-spectrum of the eligible frequencies. The true frequencies are 25Hz and 50Hz. The number of eligible frequencies are $p\cdot K=10$. We can see that the eligible frequencies all differ by ${l}_{k}\cdot f_{H}/p$ from the true frequencies.

\begin{figure}[!htbp]
\centering
\includegraphics[scale=0.8]{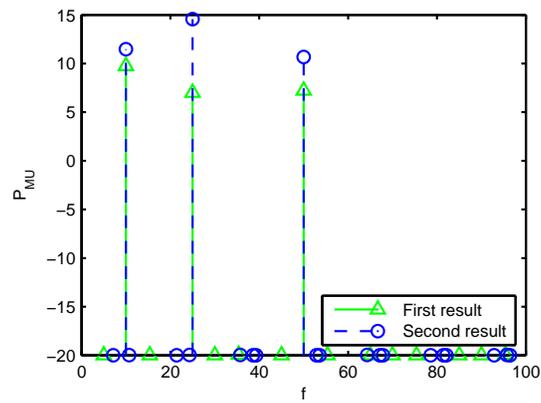}
\caption{ The pseudo-spectrum with respect to $\tilde{f}_k$ via the MUSIC-like algorithm. } \label{f3}
\end{figure}

In the second simulation, we set $K=3$ and the frequencies are set to be $10Hz,25Hz$ and $50Hz$ respectively. The white Gaussian noise at SNR=10dB is added. We use the first step of ESPRIT for the sequence at sampling rate $100/p$ and the second step of the MUSIC-like algorithm for the sequence at sampling rate $100/q$ to draw one pseudo-spectrum. Then we reverse the two sequences to draw the other. As shown in Fig.~\ref{f3}, the two pseudo-spectrums indicate the same result.

\begin{figure}[!htbp]
\centering
\includegraphics[scale=0.8]{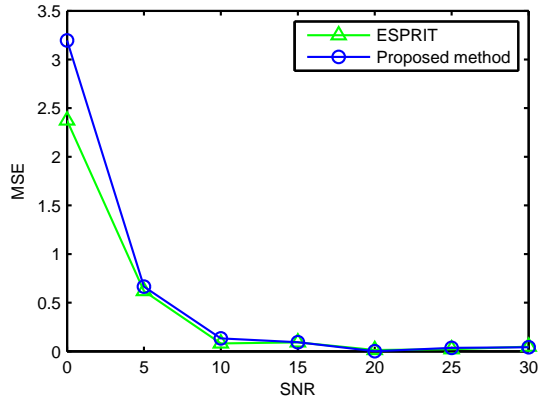}
\caption{ The comparison of errors at different SNR levels. } \label{f4}
\end{figure}

Then we compare the accuracy of the method with conventional ESPRIT. The same signals are sampled at 100Hz for the conventional ESPRIT. The number of the components is set to be 3 and the SNR varies from 0dB to 30dB. The amplitudes are set to be randomly distributed in $[0.1,1]$. The mean square error of estimated frequencies different from true frequencies is computed by $MSE = {{\sqrt {\sum\limits_{k = 1}^K {{{\left( {{{\bar f}_k} - {f_k}} \right)}^2}} } } \mathord{\left/
 {\vphantom {{\sqrt {\sum\limits_{k = 1}^K {{{\left( {{{\hat f}_k} - {f_k}} \right)}^2}} } } K}} \right.
 \kern-\nulldelimiterspace} K}$. Average values are obtained by 500 experiments for each SNR. As shown in Fig.~\ref{f4}, the proposed method mainly has the same accuracy with conventional ESPRIT, except when SNR=0dB.
\section{Conclusion}
In this paper, we propose a sub-Nyquist method to estimate the frequencies of frequency-sparse signals. Two sample sequences are sampled at two specific rates. Through the ESPRIT algorithm for one sequence and the screening algorithm for the other, the real frequencies are determined. The MUSIC-like algorithm is proven to be valid by numerical experiments. The proposed method requires less hardware than other methods and holds enough accuracy. This method can be used in many applications such as harmonic analysis, direction-of-arrival estimation, etc.
\section*{References}
\bibliography{mybibfile}

\end{document}